\documentclass[
floatfix, 
twocolumn,
showpacs, 
showkeys, 
preprintnumbers, 
nofootinbib, 
superscriptaddress
]{revtex4-1}
\usepackage[utf8]{inputenc}
\usepackage[sort&compress]{natbib}
\usepackage{ulem}
\usepackage{bm}
\usepackage{times}
\usepackage{amssymb,amsbsy,amsmath,amsfonts}
\usepackage{graphicx}
\usepackage{float}
\usepackage{color}
\usepackage{morefloats}
\usepackage{rotating}
\usepackage{srcltx}
\usepackage{slashed}
\usepackage{subfigure}
\usepackage{multirow}
\usepackage{verbatim}
\usepackage{hyperref}
\usepackage{tabularx}
\usepackage{braket}
\usepackage{lipsum}

\def\d{\mathrm{d}}

\def\l{\langle}

\DeclareUnicodeCharacter{3000}{HEREHEREHERE}

\begin{document}

\title{Neutron-deuteron scattering revisited with the EKM chiral nuclear force and the WPCD method }

\author{Qing-Yu Zhai}
\affiliation{School of Physics, Beihang University, Beijing 102206, China}
\affiliation{Department of Physics, Graduate School of Science, The University of Tokyo, Tokyo 113-0033, Japan}

\author{Dan-Yang Pang}
\affiliation{School of Physics, Beihang University, Beijing 102206, China}

\author{Wen-Di Chen}
\affiliation{Institute of Applied Physics and Computational Mathematics, Beijing, 100094, People’s Republic of China}

\author{O. A. Rubtsova}
\affiliation{Skobeltsyn Institute of Nuclear Physics, Moscow State University, 119991 Moscow, Russia}

\author{Rui-Rui Xu}
\affiliation{China Nuclear Data Center, China Institute of Atomic Energy, Beijing, 102413, China}

\author{Jun-Xu Lu}
\email[Corresponding author: ]{ljxwohool@buaa.edu.cn}
\affiliation{School of Physics, Beihang University, Beijing 102206, China}

\author{Haozhao Liang}
\email[Corresponding author: ]{haozhao.liang@phys.s.u-tokyo.ac.jp}
\affiliation{Department of Physics, Graduate School of Science, The University of Tokyo, Tokyo 113-0033, Japan}
\affiliation{Quark Nuclear Science Institute, The University of Tokyo, Tokyo 113-0033, Japan}
\affiliation{RIKEN Center for Interdisciplinary Theoretical and Mathematical Sciences, Wako 351-0198, Japan}

\author{Li-Sheng Geng}
\email[Corresponding author: ]{lisheng.geng@buaa.edu.cn}
\affiliation{Sino-French Carbon Neutrality Research Center, \'{E}cole Centrale de P\'{e}kin/School
of General Engineering, Beihang University, Beijing 100191, China}
\affiliation{School of Physics, Beihang University, Beijing 102206, China}
\affiliation{Peng Huanwu Collaborative Center for Research and Education, Beihang University, Beijing 100191, China}
\affiliation{Beijing Key Laboratory of Advanced Nuclear Materials and Physics, Beihang University, Beijing 102206, China }
\affiliation{Southern Center for Nuclear-Science Theory (SCNT), Institute of Modern Physics, Chinese Academy of Sciences, Huizhou 516000, China}

\begin{abstract}
We revisit the neutron-deuteron scattering using the Wave-Packet Continuum Discretization (WPCD) method with the EKM chiral nuclear force at various chiral orders. We rederive the permutation operator and solve the Faddeev-AGS equations directly, without rewriting the initial Faddeev kernel $tG_0$ and introducing pseudo-states, thereby rendering the approach easily extendable to a relativistic framework. We find that up to the next-to-next-to-next-to-leading order (N$^3$LO), although one can well describe the differential cross sections, one cannot resolve the long-standing $A_y$ puzzle, consistent with previous studies. The fact that the N$^3$LO chiral forces can well describe the $NN$ phase shifts and the results obtained with the EKM and Idaho N$^3$LO chiral forces agree with each other underscores the need for further investigations to resolve the $A_y$ puzzle, e.g., considering three-body forces or relativistic effects. 
\end{abstract}

\maketitle

\section{Introduction}

Neutron-deuteron ($nd$) scattering is one of the most extensively studied processes in nuclear physics. As a three-body system, it can be solved with $ab$-$initio$ methods, which serve as a standard reference for testing nuclear forces and validating few-body methods. 

In 1960, L.D. Faddeev proposed the now well-known Faddeev equations for a rigorous solution of quantum three-body problems~\cite{Faddeev:1960su}. As a set of three integral equations equivalent to the Lippmann-Schwinger equation for two-body systems, the Faddeev equations feature a compact integral kernel composed of two-body $t$ matrices and the three-body Green function. In fact, the Faddeev formalism has been generalized to more than three particles by Yakubovsky~\cite{Yakubovsky:1966ue}. Building on the Faddeev formalism, in 1967, E.O. Alt, P. Grassberger, and W. Sandhas introduced the Alt-Grassberger-Sandhas (AGS) equation, which reformulates the scattering process in terms of transition operators instead of three-body wave functions, thereby simplifying the computation of physical observables. Furthermore, H. Witala et al. developed a method to solve the three-nucleon Faddeev equations including relativistic features, in which the Green functions and permutation operators take relativistic forms and the $NN$ interaction is supplemented by boost corrections~\cite{Witala:2004pv,Witala:2008va,Witala:2011yq}. 

In general, the Faddeev-AGS equations are solved in momentum space, with a systematic method proposed in Ref.~\cite{Gloeckle:1995jg}. However, in this approach, one must deal with complicated $t$-matrix interpolations and the moving singularities of the three-body Green function, which has traditionally required the use of supercomputers. Since the $t$ matrices can be obtained analytically with separable potentials, a simplified approach is to approximate the nuclear force by an expansion in separable potentials~\cite{Ernst:1973zzb,Ernst:1974zza}, whereby the Faddeev-AGS equations can be reformulated into quasiparticle equations that are much easier to solve; however, this method is inadequate for accurately treating three-nucleon systems with high-precision nuclear forces. Another perturbative approach for solving the Faddeev-AGS equations was developed in Ref.~\cite{Deltuva:2003zp}. By taking into account one part of the interaction exactly and the other part approximately, one can decompose the two-baryon and three-baryon transition matrices into two parts, and then the Faddeev-AGS equations can be solved by iteration. Even so, one still has to deal with the complex numerical computations arising from interpolation and singularities. Recently, O.A. Rubtsova, V.I. Kukulin, and V.N. Pomerantsev developed the Wave-Packet Continuum Discretization (WPCD) method~\cite{Rubtsova:2015owa}. In this approach, the continuum states of a three-body system are coarse-grained into a square-integrable basis, thereby smoothing out all singularities and facilitating straightforward numerical solutions of the Faddeev-AGS equations. Combined with GPU-based parallel computing, this method enables efficient solution of the equations even on a personal computer~\cite{POMERANTSEV2016121}. 

In Ref.~\cite{Miller:2022beg}, the authors systematically investigated the algorithms, convergence, and other related aspects involved in solving the Faddeev-AGS equations with the Idaho next-to-next-to-next-to-leading order (N$^3$LO) and chiral optimized next-to-next-to-leading order (N$^2$LO$_\mathrm{opt}$) interactions using the WPCD method. In contrast, in the present work, we solve the Faddeev-AGS equations with the initial Faddeev kernel $tG_0$, thereby avoiding any pseudo-state constructions that rely on the Hamiltonian spectrum and can introduce theoretical inconsistencies in relativistic calculations. In addition, we propose a new equivalent expression for the permutation operators. These two technical alternatives yield the same results as the standard Faddeev calculations in standard nonrelativistic studies but are essential for generalizing to a relativistic framework, which will be the subject of future work. 

This work is organized as follows. In Sec. II, we briefly present our new formalism for solving the Faddeev-AGS equations and calculating physical observables. Results and discussions are presented in Sec. III, followed by a summary and outlook in the last section. 

\section{Theoretical Formalism}

\subsection{ Faddeev-AGS equations in momentum space }

We consider protons and neutrons as identical particles with isospin $1/2$. Therefore, elastic $nd$ scattering without three-nucleon forces is treated using the AGS equations for three identical particles. Explicitly, the AGS equations of the transition operator $U$ read
\begin{align}
    U=Pv+PtG_0U,
    \label{AGS equation}
\end{align}
where $v$ is the $NN$ interaction, $t$ is the two-body $t$-matrix defined by the Lippmann-Schwinger equation, $P=P_{12}P_{23}+P_{13}P_{23}$ is the permutation operator, which determines the overlap between different Jacobi channels, and
\begin{align}
    G_0=\frac{1}{E+i\epsilon-H_0},\ \text{with}\ H_0=E_p+E_q\equiv\frac{p^2}{m}+\frac{3q^2}{4m}\notag
\end{align}
is the free three-particle propagator, where $m$ is the mass of the nucleon and $E$ is the kinetic energy of the three-body system, $p$ and $q$ denote the standard Jacobi momenta, as illustrated in Fig.~\ref{configuration}. 

\subsection{ Setting up the WPCD basis }

We define free wave-packets using plane-wave states $|p\rangle$. Since the plane-wave basis spans an infinite-dimensional Hilbert space, to set up the WPCD method, first we need to choose a momentum cutoff $p_{\mathrm{cut}}$ and discretize continuous momenta. From the perspective of effective field theories, we are primarily concerned with the low-energy region, where the momentum grid should be chosen more densely. In this work, we use the Chebyshev grid~\cite{Rubtsova:2015owa}
\begin{align}
    p_n&=p_s\tan\left(\frac{n}{N+N_{\mathrm{add}}+1}\frac{\pi}{2}\right),\ n=0,1,2,\cdots,N,\notag\\
    p_s&=\frac{p_{\mathrm{cut}}}{\tan\left(\frac{N}{N+N_{\mathrm{add}}+1}\frac{\pi}{2}\right)},\notag
\end{align}
with $p_{\mathrm{cut}}=4.0\ \mathrm{GeV}$, $N_{\mathrm{add}}=2$, and $N=150$~\cite{Miller:2022beg}. 

Secondly, we define a set of free wave-packets as
\begin{align}
    |p_i\rangle=\frac{1}{N_i}\int_{\mathcal{D}_i}p\ \d p\ f(p)|p\rangle, \notag
\end{align}
where $N_i$ is the normalization factor and $\mathcal{D}_i\equiv[p_{i-1},p_i]$. In this work, we use the energy wave-packet, i.e., $f(p)=\sqrt{p}$ and $N_i=\sqrt{p_i^{\mathrm{mid}} \Delta p_i}$, where $\Delta p_i=p_i-p_{i-1}$ and $p_i^{\mathrm{mid}}=(p_i+p_{i-1})/2$. Further details of the WPCD method can be found in Refs.~\cite{Rubtsova:2015owa,Miller:2021vby}. 

In this paper, we label the two nucleons in the two-body subsystem as 2 and 3, and the spectator nucleon as 1. We employ the partial-wave (LS) representation, 
\begin{align}
    |p_mq_i\gamma\rangle\equiv\left|p_mq_i\left(ls\right)j\left(js_1\right)\Sigma\left(\lambda\Sigma\right)JM\left(tt_1\right)T\tau_T\right\rangle.
    \label{partial-wave basis}
\end{align}
where $p$, $l$, $s$, $j$, and $t$ denote the relative momentum, relative orbital angular momentum, spin, total angular momentum, and isospin of the antisymmetric two-body subsystem (23). The variables $q$ and $\lambda$ represent the momentum and orbital angular momentum of nucleon 1 relative to the center of mass of nucleons 2 and 3. The ``spins" of the two parts (i.e., $j$ and $s_1 = 1/2$) are coupled to form the ``total spin" $\Sigma$. The total angular momentum and total isospin of the three-body system are $J$ and $T$, respectively. In this work, we fix $T=1/2$ and $\tau_T=-1/2$. The momentum-space part of Eq.~(\ref{partial-wave basis}) is the direct product of the wave packets of the two independent Jacobi momenta, which reads
\begin{align}
    |p_mq_i\rangle=\frac{1}{\sqrt{p_m^{\mathrm{mid}} q_i^{\mathrm{mid}} \Delta p_m\Delta q_i}}\int_{\mathcal{D}_{p_m}\mathcal{D}_{q_i}} p\ \d p\ q\ \d q\ \sqrt{pq}\ |pq\rangle.\notag
\end{align}

\subsection{ Computational implementation }

Using the basis of Eq.~(\ref{partial-wave basis}), obtaining explicit matrix elements of these operators in the wave-packet representation is not always straightforward. Therefore, we next explain how to calculate the wave-packet matrix elements. 

\subsubsection{$G_0$ matrix}

First, for $G_0$, many studies applied elaborate treatments to the moving singularities~\cite{Gloeckle:1995jg}. However, within the WPCD framework, it suffices to integrate over $G_0$ alone, which can even be done analytically. In our energy wave-packet basis, it reads
\begin{align}
    &\langle p_mq_i\gamma|G_0|p_nq_j\gamma'\rangle\notag\\
    =&\frac{1}{\Delta E_{p_m}\Delta E_{q_i}}\left[(E-E_{p_m}-E_{q_i})\log\left(E-E_{p_m}-E_{q_i}\right)\right.\notag\\
     &-(E-E_{p_m}-E_{q_{i-1}})\log\left(E-E_{p_m}-E_{q_{i-1}}\right)\notag\\
     &-(E-E_{p_{m-1}}-E_{q_i})\log\left(E-E_{p_{m-1}}-E_{q_i}\right)\notag\\
     &\left.+(E-E_{p_{m-1}}-E_{q_{i-1}})\log\left(E-E_{p_{m-1}}-E_{q_{i-1}}\right)\right],\notag
\end{align}
where $\Delta E_{p_m}=E_{p_m}-E_{p_{m-1}}$, $\Delta E_{q_i}=E_{q_i}-E_{q_{i-1}}$, and $\log(x)=\log(|x|)+i\pi\Theta(-x)$. Note that we have omitted the global $\delta$-functions $\delta_{mn}\delta_{ij}\delta_{\gamma\gamma'}$, which means that $G_0$ is diagonal. 

\subsubsection{$t$ matrix}

Secondly, $t$ has singularities at the deuteron pole~\cite{Gloeckle:1995jg}
\begin{align}
    &\langle p_mq_i\gamma|t|p_nq_j\gamma'\rangle\notag\\
    =&\frac{1}{\mathcal{N}}\int p\d p\ p'\d p' \ q\d q\ \sqrt{pp'}\ t_{\gamma\gamma'}\left(p,p';E-\frac{3q^2}{4m}\right)\delta_{ij}\notag\\
    =&\frac{1}{\widetilde{\mathcal{N}}}\int p\d p\ p'\d p' \ \d E_q\ \sqrt{pp'}\ \frac{\widetilde{t}_{\gamma\gamma'}\left(p,p';E-E_q\right)}{E-E_q-E_d+i\epsilon}\delta_{ij},\notag
\end{align}
where $\gamma^{(\prime)}$ matches the $^3S_1$$-$$^3D_1$ channel and $E_d\simeq-2.2245\ \mathrm{MeV}$ is the binding energy of the deuteron. When $E-E_d\in\left[E_{q_{i-1}},E_{q_{i}}\right]$, the matrix element of $t$ has a non-trivial imaginary part, which originates from the propagating deuteron. In Refs.~\cite{Rubtsova:2015owa,Miller:2022beg}, as well as in many other studies, this difficulty was avoided by using the identity $tG_0=vG$, where $G=(E+i\epsilon-H_0-v)^{-1}$. In this way, the $t$-operator does not appear in the equations, and the scattering wave packets---namely the pseudo-states, which are eigenstates of the Hamiltonian and can be expanded by the free wave packets in the vicinity of their eigenenergies---were introduced to diagonalize $G$. However, defining a Hamiltonian in a relativistic scheme generally leads to inconsistencies, and the same issue arises when constructing scattering wave packets. This approach is difficult to generalize to the relativistic case. Therefore, we discretize the integration energy $E_q$ along a complex spectator momentum contour (SMC)~\cite{Feng:2024wyg}, that is, we use
\begin{align}
    \Gamma_{\mathrm{SMC}}\left(E_q\right)&=E_q\notag\\
    &+iV_0\left(1-e^{(E_{q_{i-1}}-E_q)/w}\right)\left(1-e^{(E_q-E_{q_i})/w}\right)\notag
\end{align}
where the parameters are set to be $V_0=w=\left(E_{q_i}-E_{q_{i-1}}\right)/2$, and $E_q\in[E_{q_{i-1}},E_{q_i}]$. It is easy to check that the value of the integral along the contour is exactly the complex conjugate of what we need. 

\subsubsection{$P$ matrix}

The explicit form of the permutation operator $P$ has been derived in numerous studies on the three-body scattering equations~\cite{Gloeckle:1995jg,Miller:2022beg}. Nevertheless, we present here a brief derivation of an equivalent expression for $P$ that allows a straightforward generalization to the relativistic framework and is numerically tractable. 

We adopt the following approach to derive the matrix elements of the permutation operator: First, we follow the approach outlined in Ref.~\cite{Chung:186421} and construct the basis of three-body partial wave states with the same coupling orders as Eq.~(\ref{partial-wave basis}). Then, we construct the permutation operator on this basis and follow Refs.~\cite{Stadler:1997iu,Wick:1962zz,McKerrell_1964} to calculate their overlap. 

\begin{figure}[hptb]
    \centering
    \includegraphics[width=8.0cm]{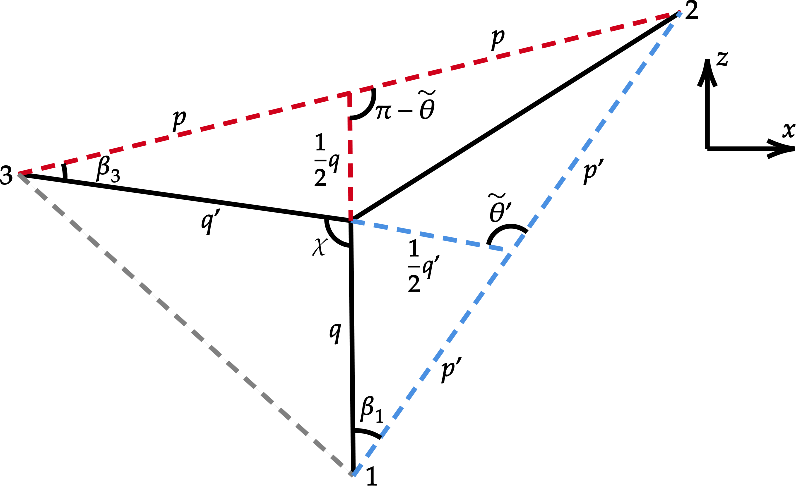}
    \caption{Momentum assignments of the three nucleons. }
    \label{configuration}
\end{figure}

Considering the partial-wave basis in Eq.~(\ref{partial-wave basis}), we illustrate the coordinate system and the relative momentum configuration of the three nucleons used in our calculation in Fig.~\ref{configuration}, where the momenta $p$ and $q$ follow the definitions in Eq.~(\ref{partial-wave basis}), whereas $p^\prime$ and $q^\prime$ represent the momenta defined in the same way, but for the configuration in which the particle labels have been permuted by $P_{13}P_{23}$. Note that we have omitted the azimuthal angle $\phi$ of the pair system, as well as the global polar and azimuthal angles $\Theta$ and $\Phi$. We denote $S=\left\{\Phi,\Theta,\phi\right\}$, which specifies the absolute orientation of the configuration depicted in Fig.~\ref{configuration} within the coordinate system.

We construct the partial-wave basis in Eq.~(\ref{partial-wave basis}) using the helicity basis $|p\lambda\rangle$, which can be easily represented by spinors and exhibits well-behaved transformation properties. We adopt the conventions of Ref.~\cite{Stadler:1997iu}
\begin{widetext}
    \begin{align}
        \mathcal{U}[L_{-z}(p)]|0-\lambda\rangle&=e^{-i\pi s}\mathcal{U}[R(\pi,\pi,0)]U[L_z(p)]|0\lambda\rangle,\notag\\
        D^{J*}_{M,m-\lambda_1}\left(\Phi,\Theta,0\right)D^{j*}_{m,\lambda_2-\lambda_3}\left(\phi,\theta,0\right)&\equiv e^{i\lambda_1\phi}D^{J*}_{M,m-\lambda_1}\left(\Phi,\Theta,\phi\right)d^{j}_{m,\lambda_2-\lambda_3}\left(\theta\right).\notag
    \end{align}
\end{widetext}
Here, $L$ denotes the transformation of changing the reference frame of a single particle state, i.e., $L_z(p)|0\lambda\rangle=|p\lambda\rangle$. Applying these relations, Eq.~(\ref{partial-wave basis}) can then be expanded in the helicity basis, leading to the explicit expressions
\begin{widetext}
    \begin{align}
        &|pq(ls)j(js_1)\Sigma(\lambda\Sigma)JM(tt_1)T\tau_T\rangle\notag\\
        =&\sum_{\substack{m,\lambda_1,\lambda_2,\lambda_3 \\ \tau_1,\tau_2,\tau_3, \tau_t}}\left(\frac{2\lambda+1}{4\pi}\right)^{1/2}\left(\frac{2l+1}{4\pi}\right)^{1/2}\l t_2\tau_2t_3\tau_3|t\tau_t\rangle\l t\tau_tt_1\tau_1|T\tau_T\rangle\notag\\
        & \quad \times \l \lambda0\Sigma m-\lambda_1|Jm-\lambda_1\rangle\l jms_1-\lambda_1|\Sigma m-\lambda_1\rangle\l l0s\lambda_2-\lambda_3|j\lambda_2-\lambda_3\rangle\l s_2\lambda_2s_3-\lambda_3|s\lambda_2-\lambda_3\rangle\notag\\
        & \quad \times \int\d\cos\theta\ \d S\ D^{J*}_{M,m-\lambda_1}\left(S\right)d^{j}_{m,\lambda_2-\lambda_3}\left(\theta\right)\mathcal{U}[R(S)]\left\{e^{-i\pi s_1}\mathcal{U}[R(\pi,\pi,0)]|q\lambda_1,t_1\tau_1\rangle\right.\notag\\
        & \qquad\quad \otimes\left.\mathcal{U}[Z(q)]\mathcal{U}[R(0,\theta,0)]|p\lambda_2,t_2\tau_2\rangle\otimes e^{-i\pi s_3}\mathcal{U}[Z(q)]\mathcal{U}[R(0,\theta,0)]\mathcal{U}[R(\pi,\pi,0)]|p\lambda_3,t_3\tau_3\rangle\right\},
        \label{h basis}
    \end{align}
\end{widetext}
where $\d S\equiv\d\cos\Theta\ \d\Phi\ \d\phi$, $R(S)=e^{-i\Phi J_z}e^{-i\Theta J_y}e^{-i\phi J_z}$, and $Z$ denotes the transformation of changing the reference frame of the pair system along the $z$-axis. $\mathcal{U}[*]$ denotes the representation of the transformation group. 

The representation of $P$ under the basis of Eq.~(\ref{h basis}) is straightforward. Note that $P_{12}P_{23}=P_{23}P_{13}P_{23}P_{23}$ and $P_{23}^{-1}=P_{23}$, thus
\begin{widetext}
    \begin{align}
        &\l pq(ls)j(js_1)\Sigma(\lambda\Sigma)JM(tt_1)T\tau_T|P_{23}P_{13}P_{23}P_{23}|p'q'(l's')j'(j's_1')\Sigma'(\lambda'\Sigma')JM(t't_1')T\tau_T\rangle\notag\\
        =&(-1)^{l+s+t}(-1)^{l'+s'+t'}\l pq(ls)j(js_1)\Sigma(\lambda\Sigma)JM(tt_1)T\tau_T|P_{13}P_{23}|p'q'(l's')j'(j's_1')\Sigma'(\lambda'\Sigma')JM(t't_1')T\tau_T\rangle\notag
    \end{align}
\end{widetext}
for nucleons, we have $l+s+t=\mathrm{odd}$, and thus $\l p_mq_i\gamma|P_{12}P_{23}|p_nq_j\gamma'\rangle=\l p_mq_i\gamma|P_{13}P_{23}|p_nq_j\gamma'\rangle$, so we only need to calculate $\l pq\gamma|P_{13}P_{32}|p'q'\gamma'\rangle$. We have
\begin{widetext}
    \begin{align}
        \l pq\gamma|P_{13}P_{23}|p'q'\gamma'\rangle=&\sum_{hh'}C^{hh'}_{\gamma\gamma'}\int\d\cos\theta\d\cos\theta'\d V\ D^{J*}_{m-\lambda_1,m'-\lambda_3'}\left(V\right)d^{j'}_{m',\lambda_1'-\lambda_2'}\left(\theta'\right)d^{j}_{m,\lambda_2-\lambda_3}\left(\theta\right)\notag\\
        &\times e^{i\pi s_1}\l q\lambda_1|\mathcal{U}^{-1}[R(\pi,\pi,0)]\mathcal{U}[R(V)]\mathcal{U}[Z(q')]\mathcal{U}[R(0,\theta',0)]|p'\lambda_1'\rangle\notag\\
        &\times e^{-i\pi s_2}\l p\lambda_2|\mathcal{U}^{-1}[R(0,\theta,0)]\mathcal{U}^{-1}[Z(q)]\mathcal{U}[R(V)]\mathcal{U}[Z(q')]\mathcal{U}[R(0,\theta',0)]\mathcal{U}[R(\pi,\pi,0)]|p'\lambda_2'\rangle\notag\\
        &\times \l p\lambda_3|\mathcal{U}^{-1}[R(\pi,\pi,0)]\mathcal{U}^{-1}[R(0,\theta,0)]\mathcal{U}^{-1}[Z(q)]\mathcal{U}[R(V)]\mathcal{U}[R(\pi,\pi,0)]|q'\lambda_3'\rangle,
        \label{P matrix element}
    \end{align}
\end{widetext}
where $h^{(\prime)}=\{m^{(\prime)},\lambda_1^{(\prime)},\lambda_2^{(\prime)},\lambda_3^{(\prime)}\}$, $R(V)\equiv R(\alpha,\chi,\beta)=R^{-1}(S)R(S')$ denotes the composited rotation, and
\begin{widetext}
    \begin{align}
        C^{hh'}_{\gamma\gamma'}=&\frac{1}{2}\frac{(-1)^t}{2J+1}\left\{\begin{array}{ccc}
        t_1 & t_2 & t' \\
        t_3 & T & t
        \end{array}\right\}\sqrt{(2l+1)(2l'+1)(2\lambda+1)(2\lambda'+1)(2t+1)(2t'+1)}\notag\\
        &\times \l \lambda0\Sigma m-\lambda_1|Jm-\lambda_1\rangle\l jms_1-\lambda_1|\Sigma m-\lambda_1\rangle\l l0s\lambda_2-\lambda_3|j\lambda_2-\lambda_3\rangle\l s_2\lambda_2s_3-\lambda_3|s\lambda_2-\lambda_3\rangle\notag\\
        &\times\l \lambda'0\Sigma'm'-\lambda_3'|Jm'-\lambda_3'\rangle\l j'm's_3-\lambda_3'|\Sigma'm'-\lambda_3'\rangle\l l'0s'\lambda_1'-\lambda_2'|j'\lambda_1'-\lambda_2'\rangle\l s_1\lambda_1's_2-\lambda_2'|s'\lambda_1'-\lambda_2'\rangle.\notag
    \end{align}
\end{widetext}

For further calculation of Eq.~(\ref{P matrix element}), in the present non-relativistic framework, all the $\mathcal{U}[Z(q^{(\prime)})]$ are the representation of the Galilean transformation, which only changes the momenta, leaving the spinor structure invariant, and the generators of the rotations are
\begin{align}
    J_y=\frac{1}{2}\begin{pmatrix}
        0 & i & 0 & 0 \\
        -i & 0 & 0 & 0 \\
        0 & 0 & 0 & i \\
        0 & 0 & -i & 0 
    \end{pmatrix},\ 
    J_z=\frac{1}{2}\begin{pmatrix}
        -1 & 0 & 0 & 0 \\
        0 & 1 & 0 & 0 \\
        0 & 0 & -1 & 0 \\
        0 & 0 & 0 & 1 
    \end{pmatrix}\notag,
\end{align}
and the helicity basis
\begin{align}
    |p\lambda\rangle\equiv\begin{pmatrix}
        1 \\ 0
    \end{pmatrix}\chi_{\lambda}\otimes|p\rangle\notag,\ 
    \text{with}\ \ \chi_{\frac{1}{2}}=\begin{pmatrix}
        1 \\ 0
    \end{pmatrix}\ \text{and}\ 
    \chi_{-\frac{1}{2}}=\begin{pmatrix}
        0 \\ 1
    \end{pmatrix}.\notag
\end{align}
According to Fig.~\ref{configuration}, we can determine the three remaining angles of rotation $V$~\cite{Stadler:1997iu}, which is $R(V)=R(0,\chi,0)$. Finally, we arrive at the expression
\begin{widetext}
    \begin{align}
        &\l p_mq_i\gamma|P_{13}P_{23}|p_nq_j\gamma'\rangle\notag\\
        =&\frac{1}{\sqrt{\Delta p_m\Delta p_n\Delta q_i\Delta q_j\ p_m^{\mathrm{mid}}p_n^{\mathrm{mid}}q_i^{\mathrm{mid}}q_j^{\mathrm{mid}}}}\sum_{hh'}\int \d p\d p'\d q\d q'\ \frac{(q)^{3/2}(q')^{3/2}}{(p)^{1/2}(p')^{1/2}}\int_0^\pi\d\chi\sin\chi\ C_{\gamma\gamma'}^{hh'}\notag\\
        &\times (-1)^{s_1-\lambda_1+\lambda_2+s_2}d^{J}_{m-\lambda_1,m'-\lambda_3'}(\chi)d^{j}_{m,\lambda_2-\lambda_3}(\widetilde{\theta})d^{j'}_{m',\lambda_1'-\lambda_2'}(\widetilde{\theta}')d_{\lambda_1\lambda_1'}^{s_1}(\beta_1)d^{s_2}_{\lambda_2\lambda_2'}(\chi-\beta_1-\beta_3)d_{\lambda_3\lambda_3'}^{s_3}(-\beta_3)\notag\\
        &\times \delta(p-\pi_1)\delta(p'-\pi_3).
        \label{P matrix element final}
    \end{align}
\end{widetext}
where 
\begin{align}
    \pi_1=\sqrt{\frac{1}{4}q^2+q'^2+qq'\cos\chi},\notag\\
    \pi_3=\sqrt{q^2+\frac{1}{4}q'^2+qq'\cos\chi},\notag
\end{align}
and the angles $\widetilde{\theta}$, $\widetilde{\theta}'$, $\beta_1$ and $\beta_3$ can be expressed by the momenta $p^{(\prime)}$ and $q^{(\prime)}$ by using the traditional law of cosines. Namely, Eq.~(\ref{P matrix element final}) is actually a triple integral of $q^{(\prime)}$ and $\chi$ without any singularities. We perform the integration using an 8-point Gaussian quadrature. 

\subsection{Spin-scattering matrix and observables}

We perform all calculations in the $NNN$ partial wave basis, treating positive and negative parity states separately, with $J\leq17/2$ and $j\leq2$\footnote{Note that Ref.~\cite{Miller:2022beg} used $j\leq3$, however, it has been proven that $j\leq2$ is sufficient to obtain accurate elastic scattering observables for $E_{\mathrm{lab}}\lesssim100\ \mathrm{MeV}$~\cite{Gloeckle:1995jg}, which is the focus of this work. }. This leads to $18$ channels for $J=1/2$, $30$ channels for $J=3/2$, and $34$ channels for $J\geq5/2$, which means that the dimension of the matrices is on the order of several hundred thousand, and direct inversion is impractical. We employ the epsilon algorithm~\cite{GRAVESMORRIS200051} to accelerate the Neumann series of the $U$-matrix, from which we obtain an approximate convergent value.

In this work, we focus on the differential cross sections and spin observable $A_y$ for elastic $nd$ scattering, which were calculated using the $6\times6$ spin-scattering matrix $\mathcal{M}$
\begin{align}
    &\mathcal{M}_{m_dm_n,m_d'm_n'}(\theta)\notag\\
    =&~\frac{2\pi i}{q_0}\sum_{J,P}\sum_{\substack{\lambda,\Sigma \\ \lambda',\Sigma'}}\sum_{\substack{m_\Sigma,m_{\Sigma'} \\ m_\lambda,M}}\l\lambda m_\lambda\Sigma m_\Sigma|JM\rangle\left\l 1m_d\frac{1}{2}m_n\middle|\Sigma m_\Sigma\right\rangle\notag\\
    &\times \l\lambda'0\Sigma'm_{\Sigma'}|JM\rangle\left\l 1m_d'\frac{1}{2}m_n'\middle|\Sigma'm_{\Sigma'}\right\rangle\notag\\
    &\times \sqrt{\frac{2\lambda'+1}{4\pi}}Y_{\lambda m_\lambda}(\theta,0)\left(S_{\lambda\Sigma,\lambda'\Sigma'}^{J^P}-\delta_{\lambda\lambda'}\delta_{\Sigma\Sigma'}\right),\notag
\end{align}
where
\begin{align}
    S_{\lambda\Sigma,\lambda'\Sigma'}^{J^P}=\delta_{\lambda\lambda'}\delta_{\Sigma\Sigma'}-i\pi\frac{4}{3}mq_0U_{\lambda\Sigma,\lambda'\Sigma'}^{J^P},\notag
\end{align}
and $q_0$ is the momentum in the c.m.s. of the $nd$ system. We can derive the observables using
\begin{align}
    \frac{\d\sigma}{\d\Omega}&=\frac{1}{6}\mathrm{Tr}\left(\mathcal{M}\mathcal{M}^\dagger\right),\notag\\
    A_y(n)&=\frac{\mathrm{Tr}\left(\mathcal{M}\Sigma_y\mathcal{M}^\dagger\right)}{\mathrm{Tr}\left(\mathcal{M}\mathcal{M}^\dagger\right)},\notag
\end{align}
where $\Sigma_y=\mathbb{I}_{3\times3}\otimes\sigma_y$. 

\section{ Numerical Results and Discussions }

To verify the validity of our framework, we compute the $nd$ scattering phase shifts for $J^P\leq7/2^\pm$ at $E_\mathrm{lab}=13\ \mathrm{MeV}$ and the neutron analyzing power $A_y(n)$ at $E_\mathrm{lab}=35\ \mathrm{MeV}$ using the Nijmegen-I and EKM $NN$ interactions, shown in Table~\ref{tab:phase shifts}. The results obtained with the Nijmegen-I potential are compared with those from a standard Faddeev calculation reported in Ref.~\cite{Gloeckle:1995jg}, serving as a benchmark. As shown in Table~\ref{tab:phase shifts}, our results are consistent with the standard Faddeev ones. For the eigen phase shifts, we reproduce the standard results with a relative error of approximately $1\%$. For the mixing parameters, the relative errors are larger while the absolute errors remain reasonable, and the impact on physical observables is limited. 

\begin{figure}[htpb]
    \centering
    \includegraphics[width=8.5cm]{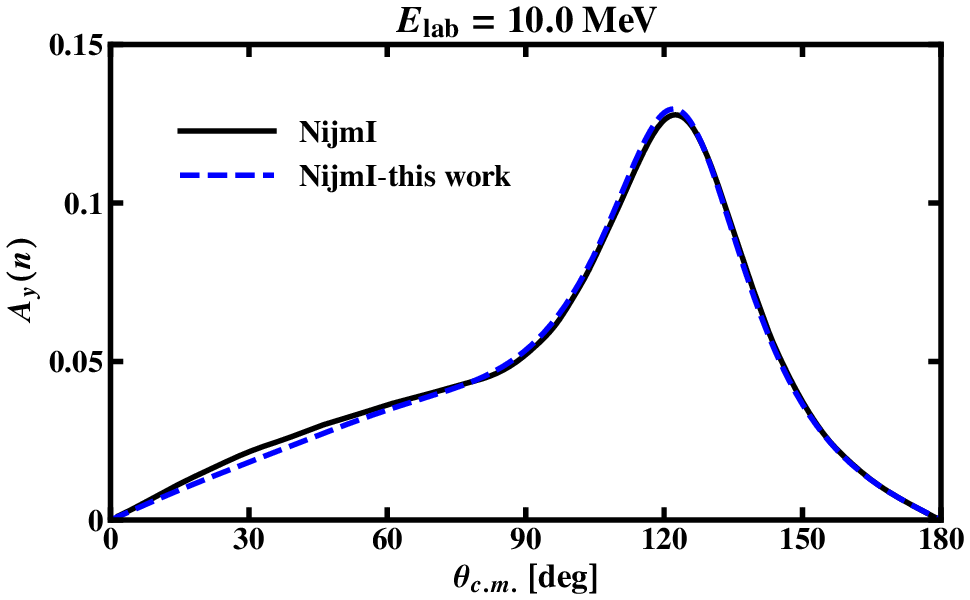}
    \caption{Neutron analyzing power $A_y(n)$ at $E_\mathrm{lab}=10\ \mathrm{MeV}$ with Nijmegen-I $NN$ interactions. The results denoted by the solid black lines are taken from Ref.~\cite{Gloeckle:1995jg}. }
    \label{comparison}
\end{figure}

A more straightforward way to benchmark our computation is by comparing observables. In Fig.~\ref{comparison}, we compare with the results from Ref.~\cite{Miller:2022beg} of the neutron analyzing power $A_y(n)$ at $E_\mathrm{lab}=10\ \mathrm{MeV}$ using the Nijmegen-I $NN$ potential. The perfect agreement verifies our approach is consistent with the standard method. Here, we would like to stress that our approach can be easily modified to suit the relativistic chiral forces recently developed~\cite{Lu:2021gsb,Lu:2025syk}. 

\begin{widetext}
    \begin{figure*}[htpb]
        \centering
        \includegraphics[width=18.0cm]{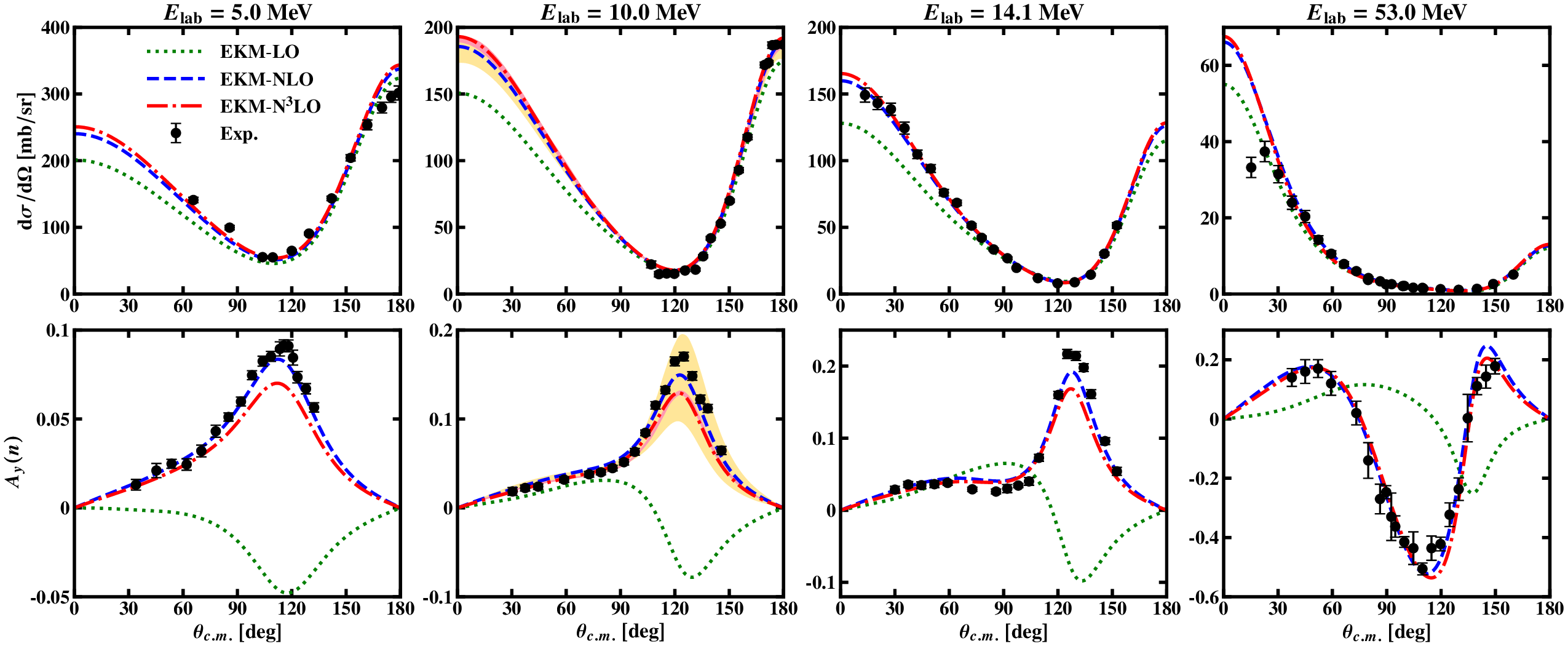}
        \caption{Differential cross sections $\d\sigma/\d\Omega$ and the neutron analyzing powers $A_y(n)$ for elastic $nd$ scattering obtained with different chiral orders of the EKM $NN$ interaction. The experimental data at $E_{\mathrm{lab}}=5$ and $10\ \mathrm{MeV}$ are taken from the EXFOR database~\cite{Otuka:2014wzu}, while the data at $E_{\mathrm{lab}}=14.1$ and $53\ \mathrm{MeV}$ are taken from Refs.~\cite{Berick:1968zz,Romero:1982zz,Watson:1982zz}. The yellow and red bands at $E_{\mathrm{lab}}=10\ \mathrm{MeV}$ are taken from Ref.~\cite{LENPIC:2015qsz}, which show the estimated theoretical uncertainties at NLO and N$^3$LO, respectively. }
        \label{results}
    \end{figure*}
\end{widetext}

We also studied the differential cross sections $\d\sigma/\d\Omega$ and the neutron analyzing power $A_y$ in the region $E_{\mathrm{lab}}\lesssim 50\ \mathrm{MeV}$ with the EKM $NN$ interaction of different chiral orders, the results of which are shown in Fig.~\ref{results}. Inspection of Fig.~\ref{results} indicates that the EKM-N$^3$LO interaction reproduces the data reasonably well, with deviations in $A_y$ around $\theta_{c.m.}\simeq120^\circ$ only for $E_\mathrm{lab}\lesssim53\ \mathrm{MeV}$, known as the $A_y$ puzzle. For the differential cross sections, different orders exhibit clear convergence, and even the lowest order provides a satisfactory description in the angular range $90^\circ\lesssim\theta\lesssim180^\circ$. In contrast, the differential cross section at small angles is more sensitive to the fine details of the nuclear force. For the neutron analyzing power, variations of different orders are strong; the lowest-order term fails to reproduce even the sign of the peak in the experimental data. On the other hand, although EKM-NLO gives a better description than EKM-N$^3$LO, it cannot give a reasonable description of the $NN$ phase shifts, thus it is not a resolution of the $A_y$ puzzle but an alternative wording of the $A_y$ puzzle. 

\begin{widetext}
    \begin{table*}[htpb]
        \centering
        \caption{Eigen phase shifts and mixing parameters for elastic $nd$ scattering at $E_\mathrm{lab}=13\ \mathrm{MeV}$. The format is (real part, imaginary part). The $\mathrm{Nijm\ I}$ results in the first column are taken from Ref.~\cite{Gloeckle:1995jg}. }\label{tab:phase shifts}
        \begin{tabular}{ c c c c c | c c c c c }
            \hline\hline
            $J^P$ & $\delta_{\Sigma\lambda}$ & $\mathrm{Nijm\ I}$ & $\mathrm{Nijm\ I}$ & $\mathrm{EKM}\left(\mathrm{N^3LO}\right)$ & $J^P$ & $\delta_{\Sigma\lambda}$ & $\mathrm{Nijm\ I}$ & $\mathrm{Nijm\ I}$ & $\mathrm{EKM}\left(\mathrm{N^3LO}\right)$ \\
            \hline
            $\frac{1}{2}^+$ & $\delta_{\frac{3}{2}2}$ & $(-7.67,0.57)$ & $(-7.52,0.58)$ & $(-7.64,0.57)$ & $\frac{1}{2}^-$ & $\delta_{\frac{1}{2}1}$ & $(-0.83,8.22)$ & $(-1.31,8.20)$ & $(-0.45,8.61)$ \\
            & $\delta_{\frac{1}{2}0}$ & $(-73.03,18.76)$ & $(-73.65,18.40)$ & $(-74.21,18.77)$ & & $\delta_{\frac{3}{2}1}$ & $(36.73,3.41)$ & $(37.81,3.53)$ & $(38.12,3.68)$ \\
            & $\eta$ & $(1.10,0.11)$ & $(1.03,0.19)$ & $(1.05,0.09)$ & & $\epsilon$ & $(22.88,4.89)$ & $(22.96,4.73)$ & $(23.10,5.14)$ \\
            $\frac{3}{2}^+$ & $\delta_{\frac{3}{2}0}$ & $(75.67,0.29)$ & $(77.62,0.62)$ & $(77.49,0.63)$ & $\frac{3}{2}^-$ & $\delta_{\frac{3}{2}3}$ & $(2.47,1.12)$ & $(2.51,1.19)$ & $(2.42,1.07)$ \\
            & $\delta_{\frac{1}{2}2}$ & $(6.84,1.56)$ & $(6.93,1.58)$ & $(6.97,1.69)$ & & $\delta_{\frac{1}{2}1}$ & $(4.82,8.52)$ & $(4.50,8.46)$ & $(5.46,9.11)$ \\
            & $\delta_{\frac{3}{2}2}$ & $(-8.15,0.57)$ & $(-8.11,0.57)$ & $(-8.12,0.57)$ & & $\delta_{\frac{3}{2}1}$ & $(30.87,2.91)$ & $(30.69,3.03)$ & $(31.50,3.09)$ \\
            & $\eta$ & $(-1.67,-0.22)$ & $(-1.76,-0.21)$ & $(-1.67,-0.22)$ & & $\eta$ & $(3.95,-19.65)$ & $(3.53,-20.58)$ & $(4.75,-18.16)$ \\
            & $\epsilon$ & $(2.03,0.51)$ & $(1.89,0.54)$ & $(2.06,0.52)$ & & $\epsilon$ & $(-15.04,-4.42)$ & $(-15.55,-4.53)$ & $(-15.02,-4.83)$ \\
            & $\xi$ & $(4.57,-0.11)$ & $(4.70,-0.11)$ & $(4.60,-0.11)$ & & $\xi$ & $(-2.24,4.94)$ & $(-2.13,5.38)$ & $(-2.46,4.52)$ \\
            $\frac{5}{2}^+$ & $\delta_{\frac{3}{2}4}$ & $(-1.00,0.02)$ & $(-0.99,0.02)$ & $(-1.00,0.02)$ & $\frac{5}{2}^-$ & $\delta_{\frac{3}{2}1}$ & $(37.55,2.41)$ & $(37.86,2.51)$ & $(38.17,2.53)$ \\
            & $\delta_{\frac{1}{2}2}$ & $(6.67,1.53)$ & $(7.36,1.36)$ & $(7.41,1.47)$ & & $\delta_{\frac{1}{2}3}$ & $(-1.18,0.22)$ & $(-1.15,0.22)$ & $(-1.17,0.24)$ \\
            & $\delta_{\frac{3}{2}2}$ & $(-9.30,0.58)$ & $(-9.34,0.58)$ & $(-9.28,0.59)$ & & $\delta_{\frac{3}{2}3}$ & $(2.99,0.08)$ & $(3.00,0.09)$ & $(3.00,0.09)$ \\
            & $\eta$ & $(-4.35,0.72)$ & $(-3.89,0.48)$ & $(-3.99,0.56)$ & & $\eta$ & $(-0.71,0.09)$ & $(-0.67,0.09)$ & $(-0.70,0.09)$ \\
            & $\epsilon$ & $(-0.62,-0.24)$ & $(-0.38,-0.26)$ & $(-0.47,-0.25)$ & & $\epsilon$ & $(0.30,0.44)$ & $(0.51,0.45)$ & $(0.31,0.46)$ \\
            & $\xi$ & $(-3.13,-0.11)$ & $(-3.25,-0.12)$ & $(-3.17,-0.12)$ & & $\xi$ & $(1.98,-0.03)$ & $(1.89,-0.03)$ & $(2.00,-0.03)$ \\
            $\frac{7}{2}^+$ & $\delta_{\frac{3}{2}2}$ & $(-7.62,0.56)$ & $(-7.57,0.57)$ & $(-7.59,0.57)$ & $\frac{7}{2}^-$ & $\delta_{\frac{3}{2}5}$ & $(0.40,0.01)$ & $(0.41,0.01)$ & $(0.40,0.01)$ \\
            & $\delta_{\frac{1}{2}4}$ & $(0.61,0.04)$ & $(0.62,0.04)$ & $(0.61,0.04)$ & & $\delta_{\frac{1}{2}3}$ & $(-1.12,0.21)$ & $(-1.10,0.21)$ & $(-1.11,0.22)$ \\
            & $\delta_{\frac{3}{2}4}$ & $(-0.98,0.02)$ & $(-0.97,0.02)$ & $(-0.98,0.02)$ & & $\delta_{\frac{3}{2}3}$ & $(3.45,0.08)$ & $(3.44,0.09)$ & $(3.47,0.09)$ \\
            & $\eta$ & $(-2.71,-0.19)$ & $(-2.81,-0.20)$ & $(-2.74,-0.20)$ & & $\eta$ & $(-9.37,-1.38)$ & $(-9.61,-1.42)$ & $(-9.47,-1.48)$ \\
            & $\epsilon$ & $(-0.45,-0.08)$ & $(-0.62,-0.08)$ & $(-0.47,-0.08)$ & & $\epsilon$ & $(0.14,-0.24)$ & $(0.07,-0.24)$ & $(0.14,-0.25)$ \\
            & $\xi$ & $(6.06,0.35)$ & $(6.29,0.38)$ & $(6.13,0.36)$ & & $\xi$ & $(-2.22,0.05)$ & $(-2.15,0.06)$ & $(-2.24,0.06)$ \\
            \hline\hline
        \end{tabular}
        \label{isospin factors}
    \end{table*}
\end{widetext}

\begin{figure}[htpb]
    \centering
    \includegraphics[width=8.5cm]{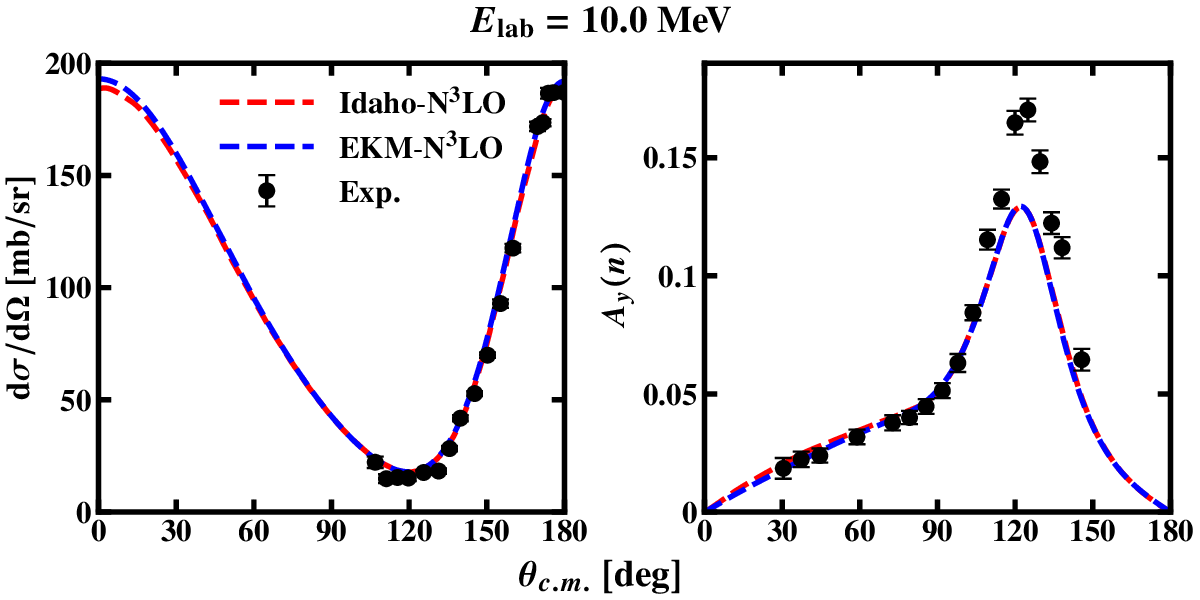}
    \caption{Differential cross section $\d\sigma/\d\Omega$ and neutron analyzing power $A_y(n)$ at $E_\mathrm{lab}=10\ \mathrm{MeV}$ with two $\mathrm{N^3LO}$ chiral $NN$ interactions. The results of $\mathrm{Idaho\text{-}N^3LO}$ are taken from Ref.~\cite{Miller:2022beg}, and the experimental data are taken from the EXFOR database~\cite{Otuka:2014wzu}. }
    \label{comparison2}
\end{figure}

In Fig.~\ref{comparison2}, we show the differential cross sections $\d\sigma/\d\Omega$ and neutron analyzing power $A_y(n)$ at $E_\mathrm{lab}=10\ \mathrm{MeV}$ with the N$^3$LO EKM and Idaho chiral $NN$ interactions. Both potentials produce essentially identical results, consistent with the fact that they can generate nearly the same $NN$ phase shifts--particularly the $^3P_j$ phases, which have a pronounced influence on $A_y$~\cite{Gloeckle:1995jg}. It is important to note that neither can reproduce the neutron analyzing power, hinting at the persistent $A_y$ puzzle, which cannot be resolved even with high-precision chiral nuclear forces, consistent with previous studies~\cite{LENPIC:2015qsz}. 

\section{ Summary and outlook }

In this work, we proposed a framework for solving the Faddeev-AGS equations using the WPCD method. While following an approach similar to Refs.~\cite{Rubtsova:2015owa,Miller:2022beg}, our study solved the Faddeev-AGS equations with the EKM chiral nuclear force directly, without introducing pseudo-states, and we presented a new equivalent expression for the permutation operator. With the two improvements above, we showed that our present framework can reproduce the results of the standard Faddeev calculation and allow a straightforward generalization to the relativistic scheme. Furthermore, we studied the differential cross sections $\d\sigma/\d\Omega$ in the region $E_{\mathrm{lab}}\lesssim 50\ \mathrm{MeV}$ with the EKM $NN$ interaction of different chiral orders and found a good reproduction of the experimental data. 

We also studied the neutron analyzing power $A_y(n)$. It was unsurprising that the $A_y$ puzzle emerged once again. Actually, this puzzle has been unresolved for over 30 years. Standard Faddeev calculations have been performed using various $NN$ interactions, and it has been shown that the theory underestimates the data by about 30\%~\cite{Gloeckle:1995jg}. It was once believed that the two-nucleon force based on chiral effective field theory (ChEFT) could resolve this puzzle; however, it persisted~\cite{Entem:2001tj}. Three-nucleon forces (3NFs) derived consistently in the framework of ChEFT have also been added to the calculation. However, it was found that adding the full N$^3$LO 3NF does not improve the description of $A_y$~\cite{Golak:2014ksa}. Moreover, the N$^3$LO 3NFs currently in use are all regularized by a multiplicative regulator applied to the 3NF expressions that are derived from dimensional regularization, which leads to a violation of chiral symmetry at N$^3$LO and destroys the consistency between two- and three-nucleon forces~\cite{Epelbaum:2019kcf}. It should be noted that the recently developed relativistic chiral nuclear force~\cite{Lu:2021gsb,Lu:2025syk} incorporates relativistic effects in a self-consistent manner and shows satisfactory convergence. It may provide new insights into this puzzle by enabling a fully relativistic study of neutron–deuteron scattering. One may expect that relativistic effects could improve the description of $A_y$, or that the puzzle might be resolved by including only the LO three-body contact terms in the relativistic power counting~\cite{Girlanda:2018xrw}.

\section{Acknowledgments}
Qing-Yu Zhai thanks Wei-Jia Kong for the useful discussions. This work is partly supported by the National Natural Science Foundation of China under Grant Nos. 12435007 and 1252200936.

\bibliography{Refs}

\end{document}